\documentclass[aps,prl,twocolumn,showpacs,preprintnumbers,nofootinbib,float,longbibliography,superscriptaddress,notitlepage]{revtex4-2}

\usepackage[T1]{fontenc}
\usepackage{lmodern}
\usepackage[utf8]{inputenc}
\usepackage{epsfig}
\usepackage{float,amsmath,amssymb}
\usepackage{graphicx,yfonts}
\usepackage{url}
\usepackage{bbold}
\usepackage{physics}
\usepackage{enumitem}
\usepackage[]{mdframed}
\usepackage[breaklinks,colorlinks,citecolor=citcolor,urlcolor=red,linkcolor=lcolor]{hyperref}
\definecolor{lcolor}{rgb}{0.5,0,0}
\definecolor{citcolor}{rgb}{0,0.3,0.0}

\begin{document}
	\title{Nonlinear collective flow reveals the breakdown of quadrupole–hexadecapole scaling in heavy ion collisions}

	\begin{abstract}
		Determining the role of intrinsic hexadecapole deformation ($\beta_4$) in nuclear structure remains a long-standing challenge. Relativistic heavy-ion collisions provide a unique opportunity to address this problem by converting the initial nuclear geometry into the collective motion of the quark--gluon plasma (QGP). Using event-by-event viscous hydrodynamic simulations of ultra-central $^{238}$U+$^{238}$U collisions at $\sqrt{s_{NN}}=193$ GeV, we investigate whether higher-order collective flow can isolate the contribution of $\beta_4$ and test the $\beta_2-\beta_4$ correlation. We demonstrate that information carried by the sign of $\beta_4$ survives the QGP evolution and is enhanced through nonlinear hydrodynamic response: the fourth-order flow harmonic acquires its topology dependence predominantly from the linear response, whereas the sensitivity of the sixth-order harmonic originates almost entirely from nonlinear mode coupling. As a consequence, the nonlinear response coefficient $\xi_{6,222}$ cleanly separates the $(\beta_2,\beta_4)$ intrinsic nuclear topologies. These results establish the sign of $\beta_4$ as an experimentally accessible signature of deviations from the quadrupole--hexadecapole correlation, demonstrating that higher-order collective flow provides a direct probe of nuclear multipole structure while revealing how nonlinear QGP dynamics encode subtle higher-order geometric information into final-state observables.
		
	\end{abstract}
	\author{Hadi Mehrabpour}
	\email{mehrabpour@fudan.edu.cn}
	\affiliation{Institute of Modern Physics, Fudan University, Shanghai 200433, China}
	\affiliation{School of Physics and Center for High Energy Physics, Peking University, Beijing 100871, China}
	\author{Zahra Sheibani}
	\affiliation{Physics Department, Yazd University, Yazd, Iran}
	\author{Li Yan}
	\email{cliyan@fudan.edu.cn}
	\affiliation{Institute of Modern Physics, Fudan University, Shanghai 200433, China}
	\affiliation{Key Laboratory of Nuclear Physics and Ion-beam Application (MOE), Fudan University, Shanghai 200433, China}
	\affiliation{Shanghai Research Center for Theoretical Nuclear Physics, NSFC and Fudan University, Shanghai 200438, China}
	\author{Chunjian Zhang}
	\email{chunjianzhang@fudan.edu.cn}
	\affiliation{Institute of Modern Physics, Fudan University, Shanghai 200433, China}
	\affiliation{Key Laboratory of Nuclear Physics and Ion-beam Application (MOE), Fudan University, Shanghai 200433, China}
	\affiliation{Shanghai Research Center for Theoretical Nuclear Physics, NSFC and Fudan University, Shanghai 200438, China}
	\author{Abolfazl Mirjalili}
	\email{a.mirjalili@yazd.ac.ir.}
	\affiliation{Physics Department, Yazd University, Yazd, Iran}
	
	\maketitle
	
	\textit{Introduction.—}
	The intrinsic shape of an atomic nucleus provides fundamental insight into the many-body dynamics of the strong interaction. Nuclear deformation is commonly characterized by multipole deformation parameters, among which the quadrupole deformation ($\beta_2$) determines the overall elongation of the nucleus, while the hexadecapole deformation ($\beta_4$) controls finer modifications of the nuclear surface. Although comparatively small, the hexadecapole deformation plays an important role in shell evolution, rotational spectra, collective excitations, and fission properties~\cite{Hendrie1968,Bemis1973,Zumbro1984,Dobaczewski2025,Guidry1976}. In particular, the sign of $\beta_4$ determines whether the nuclear density develops a waisted or barrel-like longitudinal profile, thereby defining distinct intrinsic nuclear topologies~\cite{Guidry1976,Patra:1995zz,Afanasjev:2015wwa}.
	
	Despite decades of experimental and theoretical investigations, the intrinsic hexadecapole deformation remains one of the least constrained properties of heavy nuclei. Electromagnetic transitions, Coulomb excitation, elastic scattering, and rotational spectroscopy provide only indirect access to higher-order multipole moments and suffer from significant model dependence, making the sign of $\beta_4$ particularly difficult to determine~\cite{Ryssens:2023fkv,Xu:2024bdh}. Consequently, it remains unclear to what extent the intrinsic hexadecapole deformation is constrained by the quadrupole shape through the approximate scaling relation, and whether its contribution can be experimentally isolated.
	
	In many phenomenological parameterizations and self-consistent mean-field calculations, the quadrupole and hexadecapole deformations exhibit a strong correlation, which is often approximated by the geometric scaling relation
	\begin{equation}
		\beta_4\propto\beta_2^2,
		\label{eq:scaling}
	\end{equation}
	reflecting the induced higher-order shape generated by the dominant quadrupole deformation \cite{BohrMottelson1975,RingSchuck1980,Moller1995,Moller2016,Prajapati2023,Moller:2015fba,Inakura:2025sah,4wx4-q8cj}. Eq.~(\ref{eq:scaling}) assumes that the hexadecapole deformation is largely determined by the quadrupole shape and therefore may not provide an experimentally distinguishable contribution beyond that encoded in $\beta_2$. If this approximation were valid, nuclei with identical values of $\beta_2^2$ would be expected to exhibit nearly degenerate geometric responses. Conversely, observing distinct signatures associated with the sign of $\beta_4$ would provide evidence for the breakdown of the approximate scaling relation and establish the sign of $\beta_4$ as an experimental manifestation of this violation.
	
	Ultra-relativistic heavy-ion collisions provide a fundamentally different opportunity to address this question. On time scales of order $10^{-23}$ s, The intrinsic multipole structure of the colliding nuclei is imprinted onto the initial geometry of the quark--gluon plasma~\cite{STAR:2024wgy,STAR:2025elk,Jia:2022ozr}, whose subsequent collective expansion converts this information into final-state anisotropic flow \(V_n \equiv v_n e^{in\Psi_n}\), where $v_n$ and $\Psi_n$ denote the magnitude and event-plane angle of the $n$th-order harmonic flow, respectively~\cite{Borghini:2000sa,Poskanzer:1998yz}. Unlike conventional low-energy probes, relativistic heavy-ion collisions are demonstrating that the collective response of the QGP retains detailed information about the full fluctuating nuclear geometry, making them uniquely suited to isolate signatures of higher-order multipole deformations~\cite{Mehrabpour:2026yuc,Mehrabpour:2025ogw,Mehrabpour:2023ign,Mehrabpour:2025rzt,Liu:2025uks,Bofos:2026nmg,Parida:2026uld,Giacalone:2026fat,Giacalone:2017dud,Rybczynski:2019adt,Summerfield:2021oex,Zhang:2021kxj,Xu:2021uar,Nijs:2021kvn,Zhao:2022uhl,Samanta:2023qem,Ryssens:2023fkv,Giacalone:2023cet,Fortier:2023xxy,Giacalone:2024luz,Xu:2024bdh,Zhang:2024vkh,Zhao:2024feh,Fortier:2024yxs,Giacalone:2024ixe,Mantysaari:2024uwn,Lu:2025cni,Li:2025vdp,Liu:2025zsi,Li:2025hae,3n5q-m2kf,TabatabaeeMehr:2024lgu,Taghavi:2025ddm,Li:2026igf,Zhao:2026zno,47jc-xzxm,Chen:2026gka,Liu:2025fnq,Zhao:2024lpc,Lu:2023fqd,Nielsen:2023znu,Dimri:2023wup,Magdy:2024thf}.
	
	Recent studies have established relativistic heavy-ion collisions as sensitive probes of intrinsic hexadecapole deformation. In particular, Ref.~\cite{Ryssens:2023fkv} demonstrated that a consistent implementation of surface hexadecapole deformation is essential for describing collective flow in U+U collisions, while Ref.~\cite{Xu:2024bdh} identified nonlinear response coefficients capable of determining the magnitude of $\beta_4$. In contrast, the present work addresses a different question: it is whether this information remains sufficiently robust to isolate the intrinsic hexadecapole deformation from the dominant quadrupole background.
	Rather than washing out higher-order geometric features, the collective evolution of the QGP can transfer and even amplify them through nonlinear mode coupling, making subtle aspects of the initial nuclear geometry experimentally accessible.
	
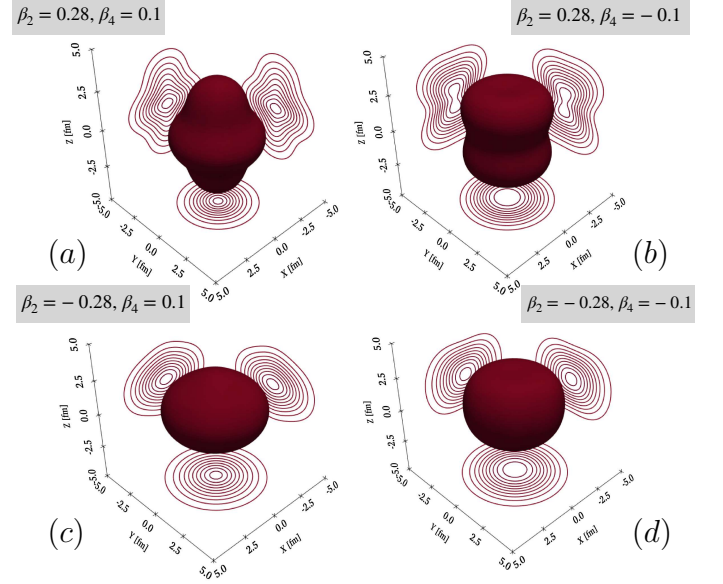
\begin{figure}[t!]
		\begin{tabular}{c} 
			\hspace*{-.5cm}\includegraphics[scale=.22]{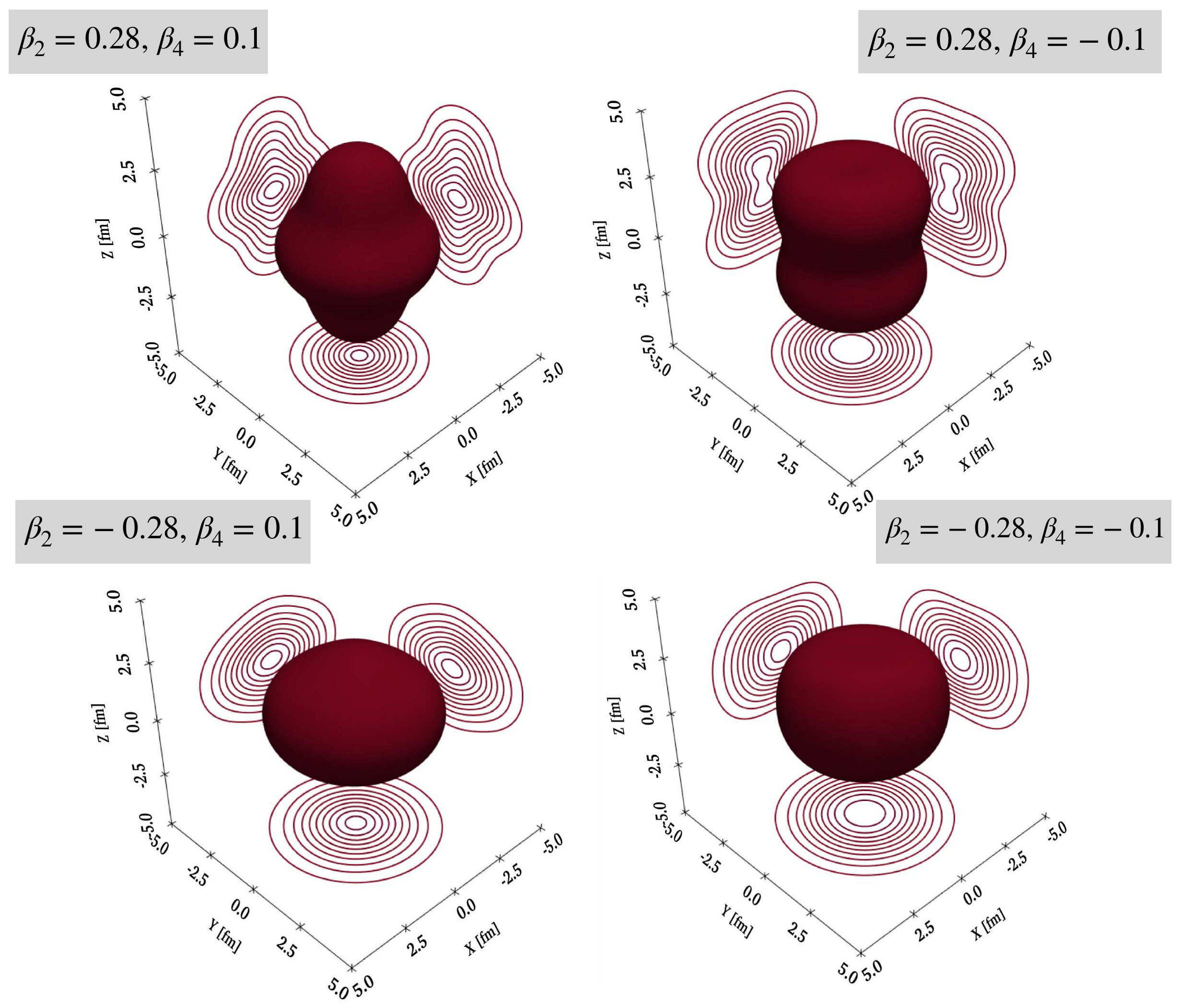}
		\end{tabular}
		\begin{picture}(0,0)
			\put(-120,125){{\fontsize{12}{12}\selectfont \textcolor{black}{$(a)$}}}
			\put(-120,20){{\fontsize{12}{12}\selectfont \textcolor{black}{$(c)$}}}
			\put(100,125){{\fontsize{12}{12}\selectfont \textcolor{black}{$(b)$}}}
			\put(100,20){{\fontsize{12}{12}\selectfont \textcolor{black}{$(d)$}}}
		\end{picture}		
		\caption{ Schematic illustration of the four intrinsic nuclear topologies considered in this work for $^{238}$U nuclei. Positive (negative) quadrupole deformation $\beta_2$ corresponds to prolate (oblate) geometry, while positive (negative) hexadecapole deformation $\beta_4$ generates waisted (barrel-like) longitudinal profiles. The four configurations shown are: (a) prolate-waisted $(\beta_2>0,\beta_4>0)$, (b) prolate-barrel $(\beta_2>0,\beta_4<0)$, (c) oblate-waisted $(\beta_2<0,\beta_4>0)$, and (d) oblate-barrel $(\beta_2<0,\beta_4<0)$. } 
		\label{fig1}
	\end{figure}

	In this work, we investigate whether the collective evolution of the quark--gluon plasma (QGP) can transfer the information associated with the intrinsic hexadecapole deformation into higher-order collective flow, thereby enabling a test of the approximate scaling relation $\beta_4\propto\beta_2^2$. Using event-by-event viscous hydrodynamic simulations of ultra-central $^{238}$U+$^{238}$U collisions, we demonstrate that the sign of $\beta_4$ emerges as a measurable signature of deviations from this scaling: the fourth-order flow harmonic acquires its topology dependence predominantly through the linear response, whereas the corresponding sensitivity of the sixth-order harmonic originates almost entirely from nonlinear mode coupling. Consequently, the nonlinear response coefficient $\xi_{6,222}$ cleanly distinguishes the four intrinsic nuclear topologies as depicted in Fig.~\ref{fig1}, revealing the breakdown of the geometric degeneracy implied by Eq.~(\ref{eq:scaling}). These findings establish higher-order collective flow as a direct probe of quadrupole--hexadecapole correlations in atomic nuclei, providing the first test of Eq.~(\ref{eq:scaling}) in relativistic heavy-ion collisions and demonstrating how nonlinear QGP dynamics transfer subtle geometric information into experimentally accessible observables.	
	
	\textit{Model setup and analysis.—}
	The central objective of this work is to determine whether higher-order collective flow preserves the approximate scaling relation $\beta_4\propto\beta_2^2$ or instead transports independent information about the intrinsic hexadecapole deformation to the final state. We therefore exploit the decomposition of the fourth- and sixth-order flow harmonics into linear and nonlinear response components \cite{Yan:2015jma,Giacalone:2018wpp},
	\begin{align}
		V_4 &= V_{4L}+\xi_{4,22}V_2^2,\nonumber\\
		V_6 &= V_{6L}+\xi_{6,222}V_2^3+\xi_{6,33}V_3^2+\xi_{6,24}V_2V_{4L},
		\label{eq:decomposition}
	\end{align}
	where $V_{nL}$ denotes the linear response to the corresponding initial eccentricity, while the remaining terms originate from nonlinear mode coupling among lower-order flow harmonics, through the nonlinear response coefficients $\xi$'s. This decomposition provides a natural framework for identifying how information associated with the intrinsic hexadecapole deformation is transported during the hydrodynamic evolution. To quantify the geometric origin of each contribution, we expand every observable around the spherical limit,
	\begin{equation}
		O=O_{\rm spherical}+\sum_{i+j\le3}\alpha_{ij}\beta_2^i\beta_4^j,
		\label{eq:expansion}
	\end{equation}
	where reflection symmetry eliminates all linear terms. If the approximate scaling relation $\beta_4\propto\beta_2^2$ were valid, the collective response would be dominated by even powers of $\beta_2$, leading to an approximate degeneracy among configurations with identical values of $\beta_2^2$. Odd powers of $\beta_4$ and mixed nonlinear terms explicitly break this degeneracy, providing the microscopic origin of the topology-dependent splitting discussed below.
	
	To quantify these effects, the intrinsic nuclear density is modeled using a deformed Woods--Saxon distribution~\cite{Miller:2007ri,PhysRev.95.577} 
	\begin{align}
		\rho(r,\theta)&=\rho_0
		(
		1+\exp\left[(r-R(\theta))/a_0\right])^{-1},
		\nonumber\\
		R(\theta)&=R_0\Big[1+\sum_{\lambda=2,4}\beta_\lambda Y_{\lambda0}(\theta)\Big],
		\label{eq:density}
	\end{align}
	where $R_0$ and $a_0$ denote the nominal nuclear radius and surface diffuseness, respectively. The deformation parameters $\beta_\lambda$ represent the multipole deformation parameters, which are directly proportional to the expectation values of the classical multipole moments $\langle \hat{Q}_\lambda \rangle = \langle r^\lambda Y_{\lambda0} \rangle$, where $Y_{\lambda m}$ are spherical harmonics.
	Although modern energy density functional (EDF) calculations provide the most microscopic description of nuclear ground-state densities, this phenomenological parameterization accurately reproduces the bulk density profiles predicted by modern energy-density-functional calculations \cite{BohrMottelson1975,RingSchuck1980,Li2024Rn224,Robledo2025History} while allowing the quadrupole ($\lambda=2$) and hexadecapole ($\lambda=4$) deformations to be varied separately, making it ideally suited for testing the validity of the approximate scaling relation in Eq. (\ref{eq:scaling}).
	
	The collision dynamics are simulated event-by-event using the iEBE--VISHNU hybrid framework with Monte Carlo Glauber initial conditions, viscous hydrodynamic evolution, and UrQMD hadronic transport~\cite{Song:2007ux,Shen:2014vra,Song:2010aq}. We study ultra-central $^{238}$U+$^{238}$U collisions at $\sqrt{s_{NN}}=193$ GeV and compare them with $^{197}$Au+$^{197}$Au collisions at $\sqrt{s_{NN}}=200$ GeV. For uranium nuclei we use $R_0=6.81$ fm, $a_0=0.55$ fm, $\beta_2=\pm0.28$, and vary the hexadecapole deformation within $-0.10\le\beta_4\le0.10$, spanning the range of phenomenological and energy-density-functional predictions. Gold nuclei are described by $R_0=6.38$ fm, $a_0=0.52$ fm, $\beta_2=-0.131$, and $\beta_4=0$. The hydrodynamic evolution starts at $\tau_0=0.6$ fm/$c$ with $\eta/s=0.08$ and switching temperature $T_{\rm sw}=160$ MeV. For each nuclear configuration, $50$k hydrodynamic events are generated in the $0$--$5\%$ centrality interval, with each event oversampled 100 times during particlization and the subsequent UrQMD evolution to ensure high statistical precision in the extracted flow observables. Charged hadrons are analyzed within $0.2<p_T<3$ GeV/$c$ and $|\eta|<1$ \cite{STAR:2024wgy}.
	
\begin{figure}[t!]
		\begin{tabular}{c} 
			\hspace*{-.5cm}
			\includegraphics[scale=.27]{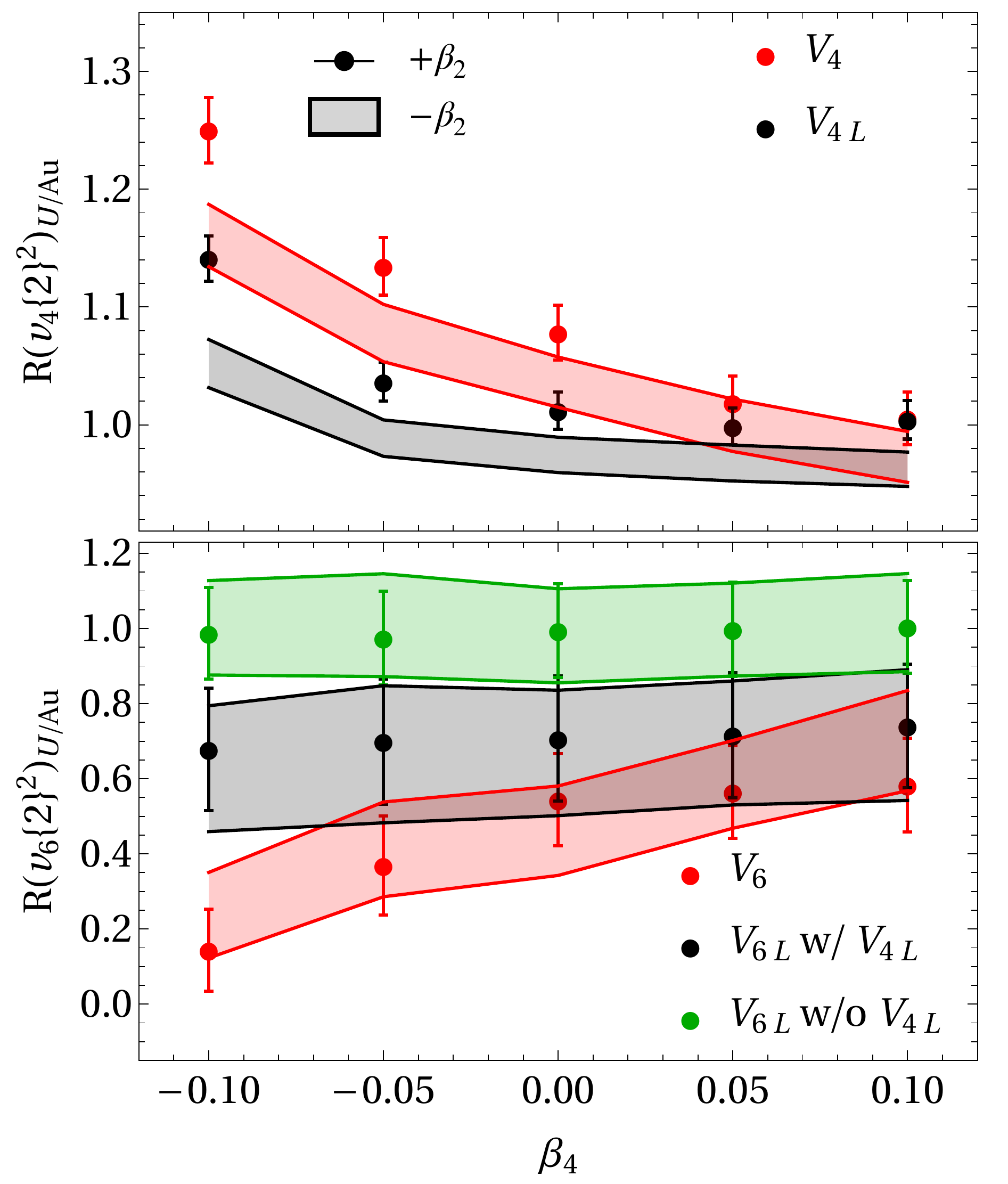}
		\end{tabular}
		\begin{picture}(0,0)
			\put(-90,90){{\fontsize{11}{11}\selectfont \textcolor{black}{iEBE-VISHNU}}}
			\put(-80,75){{\fontsize{12}{12}\selectfont \textcolor{black}{$0-5\%$}}}
			\put(-200,25){{\fontsize{12}{12}\selectfont \textcolor{black}{$(a)$}}}
			\put(-200,-105){{\fontsize{12}{12}\selectfont \textcolor{black}{$(b)$}}}
		\end{picture}		
		\caption{ Flow harmonics through two-particle correlation in ultra-central U+U collisions as functions of the hexadecapole deformation parameter $\beta_4$. Panels (a) and (b) show the ratios $R(v_4\{2\}^2)$ and $R(v_6\{2\}^2)$, respectively, for positive (markers) and negative (shade area) quadrupole deformations. The black and green markers represent the corresponding linear components extracted from the nonlinear mode-coupling decomposition. While $v_4$ exhibits enhanced sensitivity to negative $\beta_4$, the sixth-order harmonic $v_6$ develops a strong nonlinear enhancement toward positive $\beta_4$, reflecting the increasing importance of higher-order mode coupling in the collective response. }
		\label{fig2}
	\end{figure}
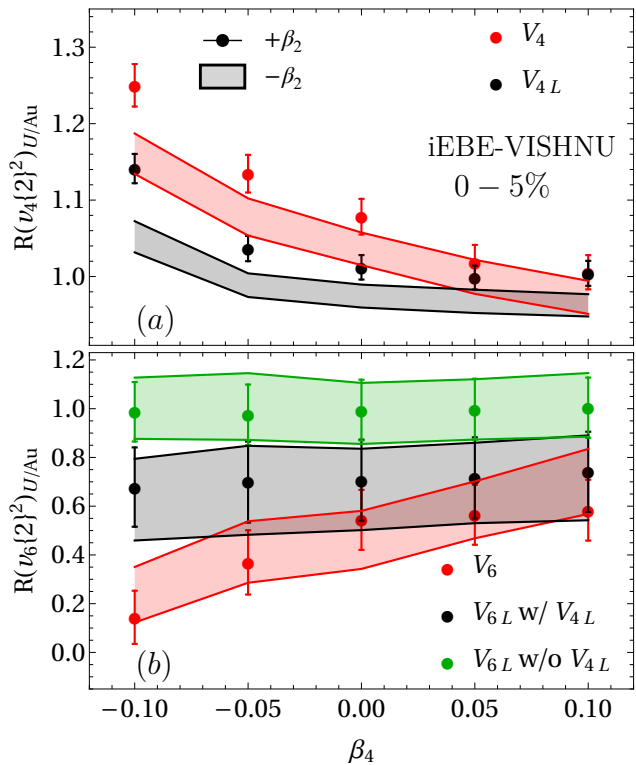

	\textit{Results and discussions.—}
	Figure~\ref{fig2} presents the ratios of the fourth- and sixth-order flow harmonics in U+U relative to Au+Au collisions as functions of the intrinsic hexadecapole deformation. The ratio construction suppresses the common hydrodynamic response and transport effects, allowing the observed splitting to be attributed predominantly to the initial nuclear geometry~\cite{Zhang:2022fou}. To identify the microscopic origin of these trends, we interpret the results using the polynomial expansion of Eq.~\ref{eq:expansion}, whose coefficients are summarized in the Supplementary Material.
	
	The ratio $R(v_4\{2\}^2)$ shown in Fig.~\ref{fig2}(a) exhibits a pronounced splitting between positive and negative values of $\beta_4$. The expansion reveals that the dominant sign-sensitive contributions originate from the cubic term $\beta_4^3$ and the mixed nonlinear coupling $\beta_2^2\beta_4$ (Eq.~\ref{aq5}), whose fitted coefficients are both negative. Consequently, negative values of $\beta_4$ produce systematically larger values of $v_4\{2\}$ than positive values with the same magnitude, leading to a clear separation between barrel- and waisted-type configurations. For $|\beta_4|=0.10$, the ratio differs by approximately $25\%$, corresponding to a statistical significance of nearly $5\sigma$, while $|\beta_4|=0.05$ still produces a measurable $\sim10\%$ ($\sim2\sigma$) separation. These results demonstrate that the sign of the intrinsic hexadecapole deformation can be experimentally distinguished through fourth-order collective flow.
	
	The behavior of $R(v_6\{2\}^2)$ in Fig.~\ref{fig2}(b) follows a qualitatively different pattern. In contrast to $v_4$, the observable is nearly insensitive to the sign of the quadrupole deformation, showing almost identical responses for prolate and oblate nuclei, while remaining highly sensitive to the sign of $\beta_4$. The expansion identifies the cubic contribution $\beta_4^3$ as the dominant sign-dependent term, whereas the mixed $\beta_2^2\beta_4$ contribution becomes negligible. As a result, the ratio increases monotonically from negative to positive $\beta_4$, indicating a positive coefficient $\alpha_{0,3}$ in Eq.~\ref{eq:expansion}. The separation between barrel- and waisted-type configurations reaches nearly $40\%$, corresponding to an experimental sensitivity of approximately $2\sigma$. The dominance of the odd-power $\beta_4$ contribution makes $v_6$ a particularly clean observable for testing the breakdown of the approximate scaling relation in Eq.~\ref{eq:scaling}.
	
	The different behavior of $v_4$ and $v_6$ originates from their distinct hydrodynamic response mechanisms. As shown in Fig.~\ref{fig2}(a) (black symbols), the topology-dependent splitting of $v_4$ is already present in its linear component $V_{4L}$, while the nonlinear contribution primarily shifts the overall magnitude through the $\beta_2^2\beta_4$ coupling (Eq.~\ref{aq7}) without altering the ordering of the four topologies. This behavior follows naturally from the linear response relation \cite{Teaney:2012ke,Teaney:2013dta}
	\begin{equation}
		V_{4L}\propto
		\mathcal{C}_4=
		\mathcal{E}_4+
		3\frac{\langle r^2\rangle^2}{\langle r^4\rangle}
		\mathcal{E}_2^2,
		\label{eq:C4}
	\end{equation}
	where the initial fourth-order eccentricity $\mathcal{E}_4$ already carries independent information on the intrinsic hexadecapole deformation, thereby breaking the approximate scaling relation of (Eq.~\ref{eq:scaling}). The sixth-order harmonic tells a complementary story. As illustrated in Fig.~\ref{fig2}(b) (black and green symbols), its linear response exhibits only weak sensitivity to the sign of $\beta_4$, whereas the nonlinear contribution generates nearly the entire topology-dependent splitting. This behavior originates from nonlinear mode coupling involving $V_4$, as demonstrated by the difference between calculations performed with and without the $V_{4L}$ contribution entering the nonlinear response. This demonstrates that the QGP encodes the sign of $\beta_4$ through two distinct response mechanisms: while the fourth-order flow is governed predominantly by the linear response to the initial geometry, the sixth-order flow acquires its topology dependence through nonlinear mode coupling.

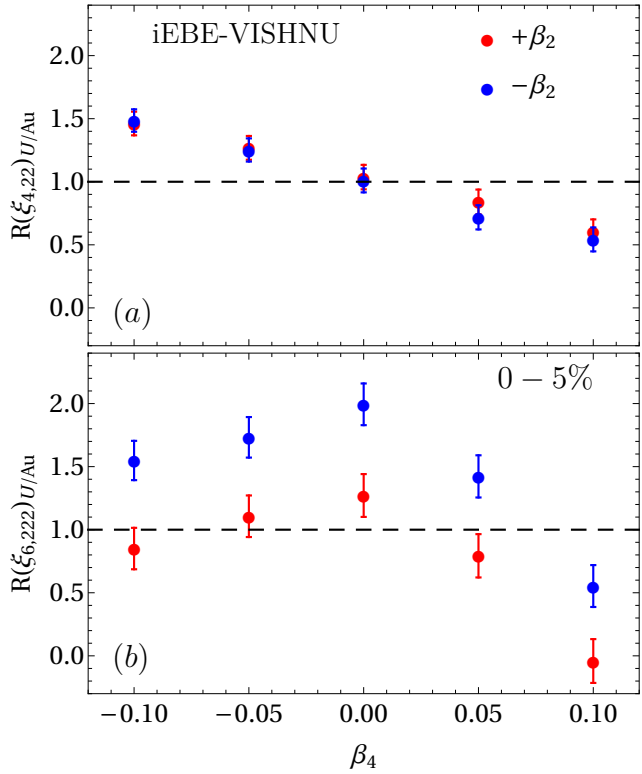
\begin{figure}[t!]
		\begin{tabular}{c} 
			\hspace*{-.3cm}\includegraphics[scale=.7]{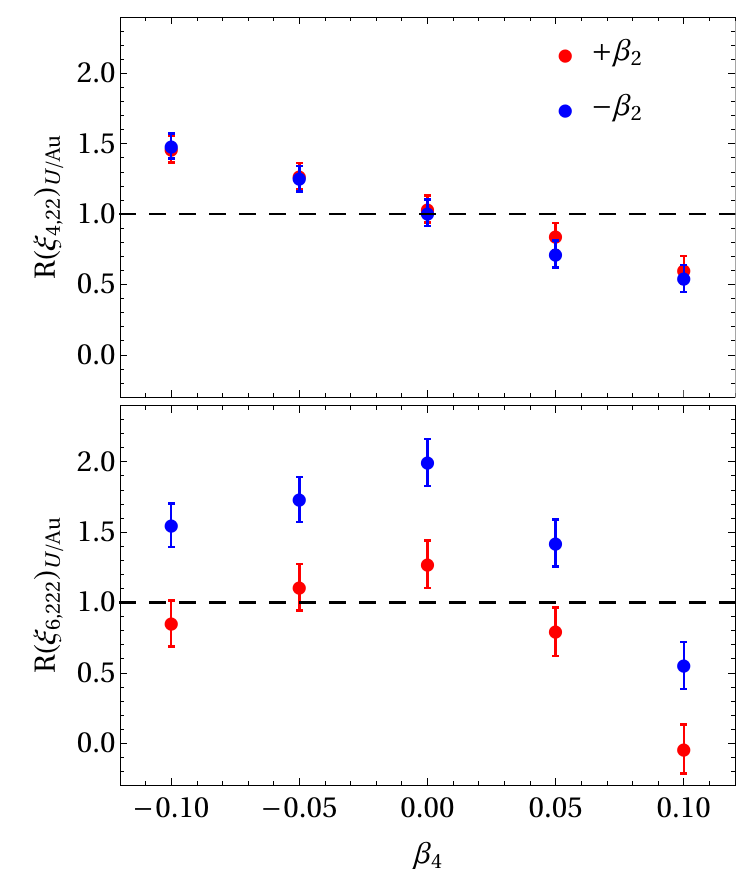}
		\end{tabular}
		\begin{picture}(0,0)
			\put(-65,278){{\fontsize{11}{11}\selectfont \textcolor{black}{iEBE-VISHNU}}}
			\put(65,148){{\fontsize{12}{12}\selectfont \textcolor{black}{$0-5\%$}}}
			\put(-80,173){{\fontsize{12}{12}\selectfont \textcolor{black}{$(a)$}}}
			\put(-80,43){{\fontsize{12}{12}\selectfont \textcolor{black}{$(b)$}}}
		\end{picture}		
		\caption{ Ratios of nonlinear response coefficients between U+U and Au+Au collisions as functions of $\beta_4$ for positive (red) and negative (blue) quadrupole deformations. Panels (a) and (b) present the nonlinear response coefficients $\xi_{4,22}$ and $\xi_{6,222}$, respectively. The fourth-order response coefficient mainly separates waisted and barrel-like configurations, whereas the sixth-order response coefficient $\xi_{6,222}$ splits all four intrinsic nuclear topologies shown in Fig.~\ref{fig1}. The large splitting observed in $\xi_{6,222}$ demonstrates the strong sensitivity of nonlinear mode coupling to the signs of both quadrupole and hexadecapole deformations. }
		\label{fig3}
	\end{figure}

	To determine how efficiently nonlinear QGP dynamics retain this information identified in Fig.~\ref{fig2}, it can be examined directly through the nonlinear response coefficients \cite{Gardim:2011xv,Giacalone:2018wpp},
	\begin{align}
		\xi_{4,22} &=
		\langle v_2^2v_4\cos(4\Psi_4-4\Psi_2)\rangle/\langle |V_2|^4\rangle,
		\label{qx4}\\
		\xi_{6,222} &=
		\langle v_2^3v_6\cos(6\Psi_6-6\Psi_2)\rangle/\langle |V_2|^6\rangle,
		\label{qx6}
	\end{align}
	which quantify the nonlinear mode coupling between $V_2$ and the higher-order collective flow. Their corresponding ratios are presented in Fig.~\ref{fig3}. The ratio $R(\xi_{4,22})$ decreases monotonically with increasing $\beta_4$, while exhibiting only a weak dependence on the sign of $\beta_2$. For $|\beta_4|=0.10$, the barrel- and waisted-type configurations differ by nearly a factor of three, corresponding to a significance exceeding $5\sigma$, whereas the separation remains about $25\%$ ($\sim2\sigma$) for $|\beta_4|=0.05$. As shown by the polynomial expansion of the corresponding observable in the Supplementary Material (Eq.~\ref{aq7}), the nonlinear response is dominated by the mixed term $\beta_2^2\beta_4$, which is the only contribution sensitive to the sign of $\beta_4$. Consequently, $\xi_{4,22}$ preserves the sign information already established by the linear response of $V_4$, rather than generating the topology-dependent ordering.
	
	A qualitatively different behavior is observed for $R(\xi_{6,222})$ in Fig.~\ref{fig3}(b). The separation between the prolate and oblate branches becomes pronounced, reaching approximately $60\%$ around $\beta_4=0$ and increasing further toward positive $\beta_4$, where the prolate branch is nearly suppressed while the oblate branch remains finite. Throughout the explored $\beta_4$ range, the two topologies are separated by more than $3\sigma$, reaching about $5\sigma$ for $|\beta_4|=0.10$. According to the expansion given in the Supplementary Material (Table.~\ref{tab1}), nearly all cubic combinations of $\beta_2$ and $\beta_4$ contribute to this observable, allowing the nonlinear mode coupling to retain simultaneous sensitivity to both deformation parameters. Consequently, $\xi_{6,222}$ resolves the four intrinsic nuclear topologies illustrated in Fig.~\ref{fig1}, establishing nonlinear mode coupling as a sensitive probe of deviations from the approximate scaling relation $\beta_4 \propto \beta_2^2$ in the sixth-order collective flow.
	
	Together, these results demonstrate that nonlinear hydrodynamic response does not merely preserve the initial hexadecapole geometry but transforms it into experimentally accessible flow correlations with enhanced topology discrimination. In particular, the sign of $\beta_4$ emerges as a measurable signature of deviations from the scaling of Eq.~(\ref{eq:scaling}), providing an experimentally accessible probe of quadrupole--hexadecapole correlations in atomic nuclei.
	
	\textit{Summary.—}
	We have investigated how intrinsic hexadecapole deformation manifests itself in higher-order collective flow in ultra-central $^{238}$U+$^{238}$U collisions. By combining event-by-event viscous hydrodynamic simulations with a systematic multipole expansion, we identified the microscopic origin of the sensitivity of fourth- and sixth-order flow harmonics to the sign of $\beta_4$. The fourth-order harmonic acquires its topology dependence predominantly through the linear response, whereas the sixth-order harmonic is governed primarily by nonlinear mode coupling.
	
	These distinct mechanisms lead to experimentally measurable signatures. The fourth-order flow harmonic discriminates barrel- and waisted-type nuclear geometries with significances reaching approximately $5\sigma$, while the nonlinear response coefficient $\xi_{6,222}$ cleanly separates the four intrinsic topologies associated with the combined signs of $\beta_2$ and $\beta_4$. The observed topology-dependent splitting originates from odd powers of $\beta_4$ and mixed quadrupole--hexadecapole couplings, providing direct evidence that higher-order collective flow is sensitive to the sign of the intrinsic hexadecapole deformation.
	
	Beyond providing a new probe of nuclear multipole structure, our results demonstrate that the QGP acts as an efficient messenger of higher-order initial-state geometry. The interplay between linear and nonlinear hydrodynamic response transfers and amplifies the information encoded in the intrinsic hexadecapole deformation, enabling experimentally accessible tests of quadrupole--hexadecapole correlations ($\beta_4\propto\beta_2^2$) in relativistic heavy-ion collisions. The present framework can be applied directly to future measurements at the RHIC and LHC and extended to other nuclei (e.g. $^{154}$Sm,  $^{150}$Nd,  $^{168}$Er,  $^{176}$Yb,  $^{208}$Pb, $^{232}$Th, etc.) with predicted hexadecapole deformations, establishing higher-order collective flow as a general probe of nuclear multipole structure.
	
	\textit{Acknowledgments.—}  This work is supported in part by the National Natural Science Foundation of China (NSFC) under Contract Nos. 12347106, 12375133, 12025501, 12547102, 12205051, the National Key Research and Development Program of China under Contract Nos. 2024YFA1612600 and 2022YFA1604900, the Natural Science Foundation of Shanghai under Contract No. 23JC1400200, Shanghai Pujiang Talents Program under Contract No. 24PJA009. A.M  and Z.S acknowledge Yazd University for providing the facilities to conduct this work.

	\bibliography{biblio}
	
	\appendix
	
	\section*{Supplementary Materials}
	\subsection*{Estimations of flow moments and cumulants}
	\begin{figure*}[t!]
		\begin{tabular}{c}
			\hspace*{-.6cm}\includegraphics[scale=.29]{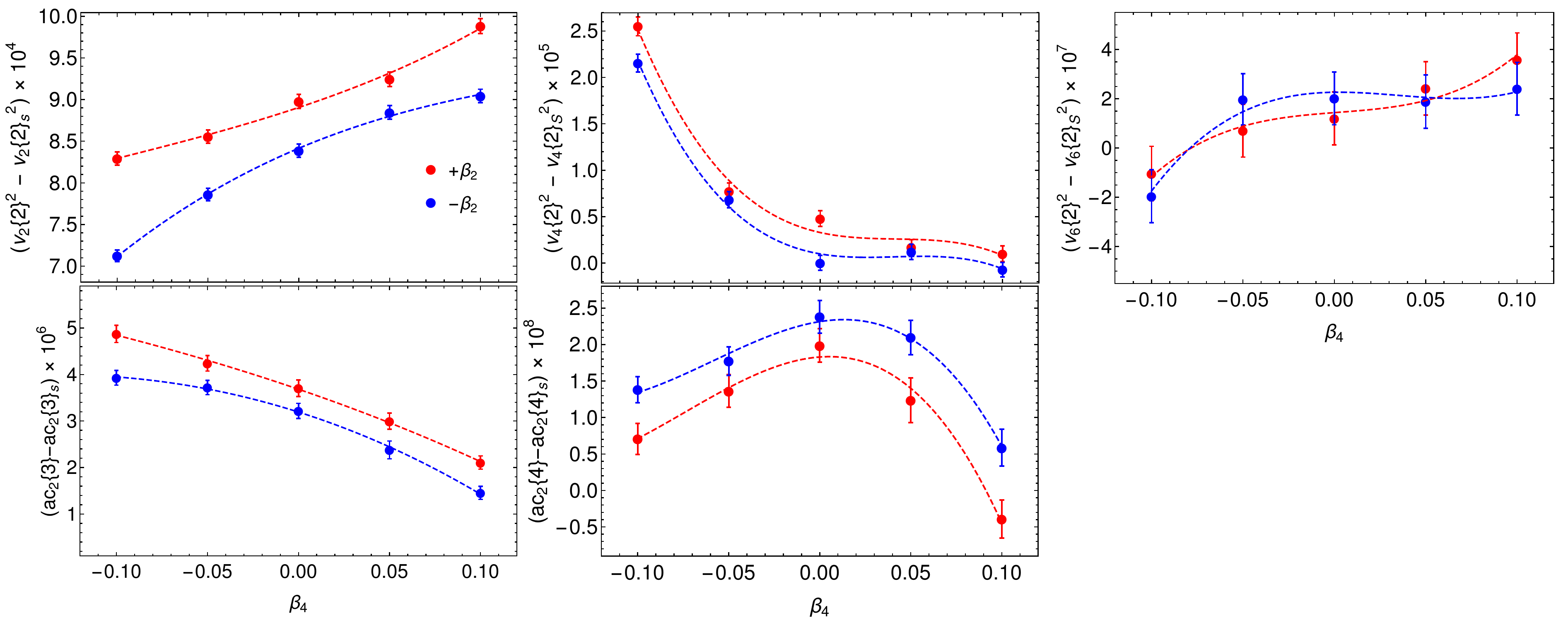}
		\end{tabular}
		\begin{picture}(0,0)
			\put(110,50){{\fontsize{12}{12}\selectfont \textcolor{black}{U+U @ $\sqrt{s_{NN}}=193$ GeV}}}
			\put(110,65){{\fontsize{12}{12}\selectfont \textcolor{black}{iEBE-VISHNU}}}
			\put(110,35){{\fontsize{12}{12}\selectfont \textcolor{black}{$0-5\%$}}}
			\put(-200,125){{\fontsize{12}{12}\selectfont \textcolor{black}{$(a)$}}}
			\put(-30,125){{\fontsize{12}{12}\selectfont \textcolor{black}{$(b)$}}}
			\put(140,125){{\fontsize{12}{12}\selectfont \textcolor{black}{$(c)$}}}
			\put(-200,33){{\fontsize{12}{12}\selectfont \textcolor{black}{$(d)$}}}
			\put(-30,33){{\fontsize{12}{12}\selectfont \textcolor{black}{$(e)$}}}
		\end{picture}		
		\caption{ Deformation-induced contributions to flow harmonics and mixed-harmonic correlations in ultra-central U+U collisions obtained within the iEBE-VISHNU framework. The panels show the deformation-dependent components of (a) $v_2\{2\}^2$, (b) $v_4\{2\}^2$, (c) $v_6\{2\}^2$, (d) $\mathrm{ac}_{2}\{3\}$, and (e) $\mathrm{ac}_{2}\{4\}$ as functions of the hexadecapole deformation parameter $\beta_4$ for positive (red) and negative (blue) quadrupole deformation. Markers represent the full hydrodynamic calculations, while dashed curves correspond to the truncated expansion of Eq.~(\ref{eq:expansion}). The progressively enhanced splitting observed for higher-order mixed-harmonic observables reflects the increasing role of nonlinear mode coupling in amplifying the underlying quadrupole--hexadecapole structure of the initial geometry. }
		\label{figapp}
	\end{figure*}
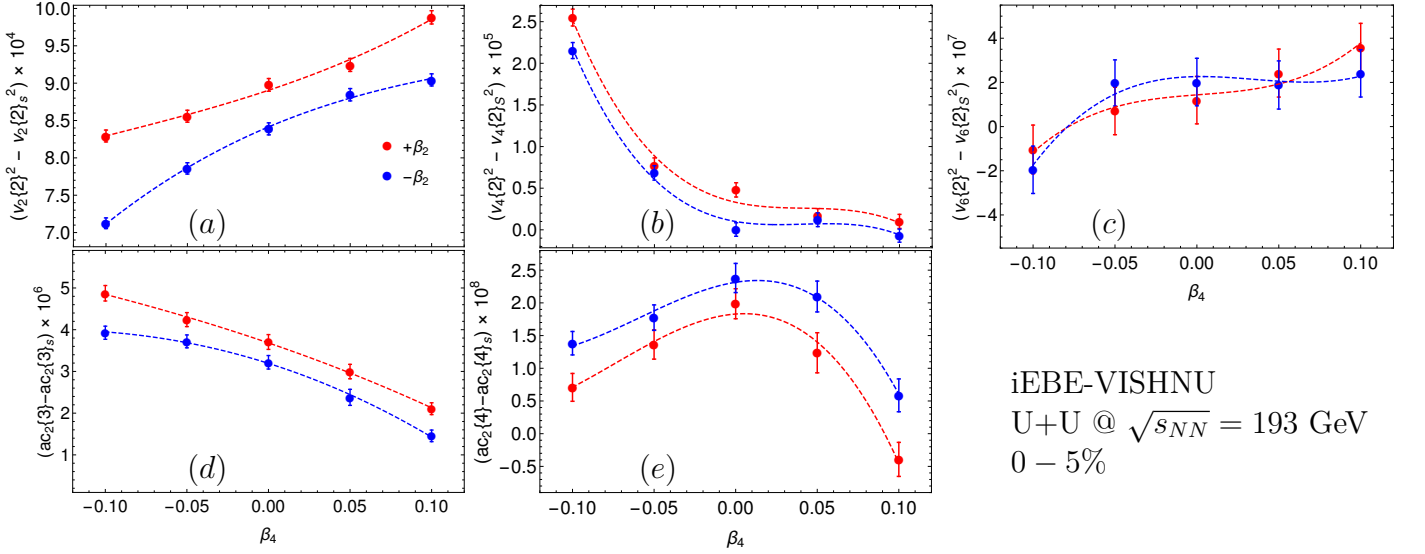
	To further investigate the interplay between quadrupole and hexadecapole deformations, we perform a complementary analysis based on a truncated expansion of the flow observables in terms of the deformation parameters $\beta_2$ and $\beta_4$ introduced in Eq.~\ref{eq:expansion}. This expansion is truncated at cubic order in order to retain sensitivity to the signs of the deformation parameters while neglecting higher-order terms with numerically negligible contributions. We further observe that the mixed linear term $\beta_2\beta_4$ is approximately consistent with zero.
	
	The coefficients $\alpha_{i,j}$ are extracted by fitting Eq.~\eqref{eq:expansion} to the observables shown by the dashed curves in Fig.~\ref{figapp}. To isolate deformation-driven contributions, the figure displays $\mathcal{O}-\mathcal{O}_s$. Since the dominant sensitivity to quadrupole--hexadecapole coupling arises from observables involving $v_2$, $v_4$, and $v_6$, we investigate the contribution of each term in Eq.~\eqref{eq:expansion} using $v_n\{2\}^2$ for $n=2,4,6$ together with the mixed-harmonic correlations
	\begin{align}
		\text{ac}_{2}\{3\}
		&\equiv
		\langle v_2^2 v_4 \cos(4\Psi_4 - 4\Psi_2)\rangle,
		\\
		\text{ac}_{2}\{4\}
		&\equiv
		\langle v_2^3 v_6 \cos(6\Psi_6 - 6\Psi_2)\rangle.
	\end{align}
	The corresponding contributions of the coefficients $\alpha_{i,j}$ are summarized in Table~\ref{tab1}.
	
	As shown in Fig.~\ref{figapp}(a), deformation part of $v_2\{2\}^2$ already exhibits clear sensitivity to the four topology classes introduced in Fig.~\ref{fig1}. A pronounced dependence on $\beta_4$ is observed, particularly for positive $\beta_2$, where quadrupole and hexadecapole deformations reinforce each other and enhance the elliptic anisotropy. Quantitatively, we obtain
	$(\text{case a}/\text{case b})-1\approx11.1\%$,
	$(\text{case c}/\text{case d})-1\approx14.5\%$,
	$(\text{case a}/\text{case c})-1\approx5.1\%$, and
	$(\text{case b}/\text{case d})-1\approx8.4\%$, where 
	\begin{itemize}
		\item case a: Prolate-waisted ($\beta_2 = +0.28, \beta_4=0.1$)
		\item case b: Prolate-barrel ($\beta_2 = +0.28, \beta_4=-0.1$)
		\item case c: Oblate-waisted ($\beta_2 = -0.28, \beta_4=0.1$)
		\item case d: Oblate-barrel ($\beta_2 = -0.28, \beta_4=-0.1$)
	\end{itemize}
	The sign of $\beta_2$ generates an additional $\sim4.5\%$ variation in $v_2\{2\}^2$, which becomes more pronounced in the ratio $(\beta_4\neq0)/(\beta_4=0)$. In particular, the difference between $(\beta_2=0.28,\beta_4=0.1)$ and $(\beta_2=0.28,\beta_4=0)$ is approximately $6\%$, consistent with the results reported in Refs.~\cite{Xu:2024bdh,Xu:2025cgx}.
	
	The decomposition of $v_2\{2\}^2$ is obtained as
	\begin{align}\label{aq4}
		v_2\{2\}^2-v_{2,s}\{2\}^2
		=
		&\alpha_{2,0}\beta_2^2
		+\alpha_{3,0}\beta_2^3
		\nonumber\\
		&
		+\alpha_{2,1}\beta_2^2\beta_4
		+\alpha_{1,2}\beta_2\beta_4^2,
	\end{align}
	where the dominant contribution originates from the expected $\beta_2^2$ term. The splitting between prolate and oblate configurations is primarily controlled by the $\beta_2^3$ contribution and, at larger $\beta_4$, by the $\beta_2\beta_4^2$ term. We further observe that the $\beta_2\beta_4^2$ contribution appears only in $v_2\{2\}^2$. In contrast, the term $\beta_2^2\beta_4$ efficiently separates waisted and barrel-like geometries, demonstrating that $v_2\{2\}$ already acts as a sensitive probe of quadrupole--hexadecapole coupling.
	
	For $v_4\{2\}^2$, the decomposition becomes
	\begin{align}\label{aq5}
		v_4\{2\}^2-v_{4,s}\{2\}^2
		=
		&\alpha_{2,0}\beta_2^2
		+\alpha_{0,2}\beta_4^2
		\nonumber\\
		&
		+\alpha_{3,0}\beta_2^3
		+\alpha_{0,3}\beta_4^3
		+\alpha_{2,1}\beta_2^2\beta_4.
	\end{align}
	The dominant contribution arises from the $\beta_4^2$ term, while the $\alpha_{2,0}\beta_2^2$ component produces a sizable overall shift. The splitting between positive and negative $\beta_2$ configurations is mainly driven by the $\beta_2^3$ contribution, although the effect remains moderate. The combined action of the $\beta_4^3$ and $\beta_2^2\beta_4$ terms generates the strong separation between waisted and barrel-like structures. In particular, the $\beta_2^2\beta_4$ contribution is responsible for maintaining
	$v_4\{2\}^2|_{\beta_4=0}>
	v_4\{2\}^2|_{\beta_4>0}$.
	
	The structure of $v_6\{2\}^2$ differs qualitatively from the lower-order harmonics,
	\begin{align}\label{aq6}
		v_6\{2\}^2-v_{6,s}\{2\}^2
		=
		\alpha_{2,0}\beta_2^2
		+\alpha_{0,2}\beta_4^2
		+\alpha_{0,3}\beta_4^3,
	\end{align}
	with no visible contribution from either $\beta_2^2\beta_4$ or $\beta_2\beta_4^2$. The dominant sensitivity originates from the $\beta_2^2$ and $\beta_4^2$ terms, while the distinction between positive and negative hexadecapole deformation is primarily generated by the cubic $\beta_4^3$ contribution. Consequently, $v_6\{2\}$ provides a particularly clean probe of the sign of $\beta_4$, although it exhibits weaker sensitivity to direct quadrupole--hexadecapole coupling.
	
	\begin{table}[b!]
		\begin{tabular}{cccccc}
			\hline
			coefficient & $v_2\{2\}$ & $v_4\{2\}$ & $v_6\{2\}$ & $\text{ac}_{2}\{3\}$ & $\text{ac}_{2}\{4\}$ \\
			\hline
			$\alpha_{2,0}$ & $\checkmark$ & $\checkmark$ & $\checkmark$ & $\checkmark$ & $\checkmark$ \\
			$\alpha_{0,2}$ & $\times$ & $\checkmark$ & $\checkmark$ & $\times$ & $\checkmark$ \\
			$\alpha_{3,0}$ & $\checkmark$ & $\checkmark$ & $\times$ & $\checkmark$ & $\checkmark$ \\
			$\alpha_{0,3}$ & $\times$ & $\checkmark$ & $\checkmark$ & $\times$ & $\checkmark$ \\
			$\alpha_{2,1}$ & $\checkmark$ & $\checkmark$ & $\times$ & $\checkmark$ & $\checkmark$ \\
			$\alpha_{1,2}$ & $\checkmark$ & $\times$ & $\times$ & $\times$ & $\times$ \\
			\hline
		\end{tabular}
		\caption{
			Contributions of the coefficients $\alpha_{i,j}$ governing the $\beta_2$ and $\beta_4$ dependence of the observables shown in Fig.~\ref{figapp} for $0$--$5\%$ U+U collisions within the iEBE-VISHNU framework.
		}
		\label{tab1}
	\end{table}
	
	The information of nonlinear modes are encoded in asymmetric cumulants \cite{ATLAS:2014qaj,Zhao:2022uhl,Jia:2017hbm}.  The results for the asymmetric cumulants are presented in panels (d) and (e) of Fig.~\ref{figapp}. Similar to $v_2\{2\}$, the correlator $\text{ac}\{3\}$ directly reflects quadrupole--hexadecapole coupling,
	\begin{align}\label{aq7}
		\text{ac}\{3\}^2-\text{ac}_{s}\{3\}^2
		=
		\alpha_{2,0}\beta_2^2
		+\alpha_{3,0}\beta_2^3
		+\alpha_{2,1}\beta_2^2\beta_4.
	\end{align}
	However, the observable mainly isolates the $\beta_2^2\beta_4$ contribution, which explains why the nonlinear response coefficient $\xi_{4,22}$ provides a more efficient discriminator of the sign of $\beta_4$, as discussed in the main text.
	In contrast, all terms except $\beta_2\beta_4^2$ contribute to $\text{ac}\{4\}$, making it a complementary probe to $v_6\{2\}$. This hierarchy explains why the nonlinear response coefficient $\xi_{6,222}$ emerges as the most powerful observable for discriminating the four topology classes illustrated in Fig.~\ref{fig1}.

\end{document}